\documentclass[superscriptaddress,prd]{revtex4}
\usepackage{epsfig}

\begin{document}
\date{\today}
\title{Constant Crunch Coordinates for Black Hole Simulations}

\author{Adrian P. Gentle}
\email{apg@lanl.gov}
\affiliation{Theoretical Division (T-6, MS B288), Los Alamos National
Laboratory, Los Alamos, NM 87544}
%\address{Theoretical Division (T-6, MS B288), Los Alamos National
%Laboratory, Los Alamos, NM 87544}

\author{Daniel E. Holz}
\email{deholz@itp.ucsb.edu}
\affiliation{Institute for Theoretical Physics, University of
California, Santa Barbara, CA 93106}
%\address{Institute for Theoretical Physics, University of
%California, Santa Barbara, CA 93106}

\author{Arkady Kheyfets}
\email{kheyfets@math.ncsu.edu}
\affiliation{Department of Mathematics, North Carolina State
University, Raleigh, NC 27695-8205}
%\address{Department of Mathematics, North Carolina State
%University, Raleigh, NC 27695-8205}

\author{Pablo~Laguna}
\email{pablo@astro.psu.edu}
\affiliation{Department of Astronomy and Astrophysics and Center for
Gravitational Physics and Geometry, Penn State University, State
College, PA 16802}
%\address{Department of Astronomy and Astrophysics and Center for
%Gravitational Physics and Geometry, Penn State University, State
%College, PA 16802}

\author{Warner A. Miller}
\email{wam@lanl.gov}
\affiliation{Theoretical Division (T-6, MS B288), Los Alamos National
Laboratory, Los Alamos, NM 87544}
%\address{Theoretical Division (T-6, MS B288), Los Alamos National
%Laboratory, Los Alamos, NM 87544}

\author{Deirdre M. Shoemaker}
\email{deirdre@astro.psu.edu}
\affiliation{Department of Astronomy and Astrophysics and Center for
Gravitational Physics and Geometry, Penn State University, State
College, PA 16802}
%\address{Department of Astronomy and Astrophysics and Center for
%Gravitational Physics and Geometry, Penn State University, State
%College, PA 16802}

%\maketitle

\begin{abstract}

We reinvestigate the utility of time-independent constant mean
curvature foliations for the numerical simulation of a single
spherically-symmetric black hole. Each spacelike hypersurface of such
a foliation is endowed with the same constant value of the trace of
the extrinsic curvature tensor, $K$.  Of the three families of
$K$-constant surfaces possible (classified according to their
asymptotic behaviors), we single out a sub-family of
singularity-avoiding surfaces that may be particularly useful, and
provide an analytic expression for the closest approach such surfaces
make to the singularity.  We then utilize a non-zero shift
to yield families of $K$-constant surfaces which (1) avoid the black
hole singularity, and thus the need to excise the singularity, (2) are
asymptotically null, aiding in gravity wave extraction, (3) cover the
physically relevant part of the spacetime, (4) are well behaved
(regular) across the horizon, and (5) are static under evolution, and
therefore have no ``grid stretching/sucking'' pathologies. Preliminary
numerical runs demonstrate that we can stably evolve a single
spherically-symmetric static black hole using this foliation.  We wish
to emphasize that this coordinatization produces $K$-constant surfaces
for a single black hole spacetime that are regular, static and stable
throughout their evolution.
\end{abstract}


\maketitle

\section[] {Constant Crunch Surfaces}
\label{sec:0}

In this paper, we address a single question: Is there a
numerically-viable coordinatization of a Schwarzschild black hole
spacetime foliated by hypersurfaces of constant (not necessarily zero)
mean extrinsic curvature?  In other words, can we coordinatize the
Schwarzschild spacetime with constant mean extrinsic curvature
($Tr(K)=\mbox{\em constant}$) hypersurfaces so as to bound the growth
of metric components and their gradients?  We demonstrate here that
the single shift freedom yields a spacetime metric that is static, and
therefore bounds the growth in time of such gradients. A more complete
analysis of the stability of our coordinatization, and a more thorough
canvassing of the parameter space, will appear
elsewhere.~\cite{Shoemaker2000} Our foliation is consistent with that
of Iriondo {\em et al.}~\cite{Iriondo1996}, who provided a generic
constant mean curvature (CMC) foliation of the Reissner-Nordstr\"om
spacetime for the purpose of finding trapped surfaces. In this paper
we focus on the utility of CMC slicings for the numerical simulation
of black holes, in support of the emerging field of gravity-wave
astrophysics.

The trace of the extrinsic curvature tensor ($Tr({\bf K})=K^a{}_a=K$)
at a point on a spacelike hypersurface measures the fractional rate of
contraction of 3-volume along a unit normal to the surface.  It
represents the amount of ``crunch'' the 3-surface is experiencing at
the point, at a given time.  If all the observers throughout a
spacelike hypersurface moving in time orthogonal to the surface
experience the same amount of contraction per unit proper time, we say
that the surface is a $K$-surface or a ``constant crunch'' surface. In
this paper we examine foliations of a single spherically-symmetric,
static black hole where each spacelike hypersurface has the same
constant value of the extrinsic curvature, $K$.

Generic $K$-surface foliations have found great utility in the
numerical simulation of cosmological spacetimes.~\cite{Centrella1984}
In addition to decoupling the three momentum constraint equations from
the Hamiltonian constraint, these surfaces (in the case of compact or
W-model universes) provide a convenient cosmological time parameter
($K$, or York, time).~\cite{Wheeler1988} Furthermore, for such
cosmological spacetimes one has powerful existence and uniqueness
theorems.~\cite{Tipler1980,Tipler1985} Extensive work into the
characteristics of these surfaces for Schwarzschild spacetimes has
been done by Brill {\em et al.},~\cite{Brill1978} and foundational work
into their use in numerical relativity was done by D. Eardley {\em et
al.}.~\cite{Eardley1978} More recently Pervez {\em et
al.}~\cite{Pervez1995} provided a foliation partially covering the
Schwarzschild spacetime with $K$-surfaces, with $K$ ranging from
$-\infty$ to $\infty$, and Iriondo {\em et al.}~\cite{Iriondo1996}
provided a generic constant mean curvature foliation of the
Reissner-Nordstr\"om spacetime for the purpose of finding trapped
surfaces. In this paper we build upon the work of these investigators
by examining the utility of these surfaces for numerical relativity in
support of gravity-wave detectors.

Although surfaces of constant $K$ were thoroughly investigated decades
ago, their use in current numerical simulations of black holes is
conspicuously absent (apart from the use of maximal ($K=0$)
surfaces).~\cite{Alcubierre2000} One reason why these slicing methods
have not been more fully developed is that they lag in time close in,
to avoid crashing into the singularity, while they simultaneously
evolve forward normally at the outer edge of the grid to allow for
wave extraction.  This tension, many fear, will unavoidingly lead to
unbounded growth in the metric and extrinsic curvature components in
the intermediate region, as is indeed found in maximal slicing.  This
computational concern has been referred to by the numerical relativity
community as ``grid stretching'' or ``grid sucking.'' We show in this
paper that a proper choice of radial shift can yield a constant crunch
foliation of a spherically-symmetric black hole without such
pathologies. In fact, we foliate a Schwarzschild black hole such that
the 3-metric and extrinsic curvature are both bounded and static
(i.e. unchanging in time).

To numerically evolve a black hole 3-space in time it is desirable to
have a foliation, and its coordinatization, which satisfy the
following four properties:
\begin{enumerate}
\item   Avoids black hole singularities or facilitates their excision.
\item   Possesses asymptotically null hypersurfaces to aid
in radiation extraction.
\item   Minimizes steep gradients in the lapse, shift, 3-metric and
        extrinsic curvature tensor.
\item   Maximizes the future development of the initial data for the purpose of
        gravity-wave extraction.
\end{enumerate}
As a first step towards achieving these goals for systems containing
multiple black holes, we explore the families of $K$-surfaces in the
Schwarzschild spacetime, and find a CMC foliation 
satisfying the above properties. 

In the next section we construct the $K$-surfaces outside and inside
the horizon. In section~\ref{sec:3} we explore the properties of the
$K$-surfaces, dwelling in particular on their approaches to the
singularity. We also examine and illustrate the three families of
$K$-surfaces. In section~\ref{sec:4} we derive a metric for
Schwarzschild whose constant time slices are
$K$-surfaces. We
 restrict our attention to a subfamily of $K$-surfaces --
surfaces which, when generalized to the colliding black hole
spacetimes, support the gravity wave detection problem. We also
present some preliminary numerical simulations using
constant-crunch coordinates. We conclude with general comments on
the applicability of $K$-surfaces to numerical calculations of more
general black hole spacetimes.

\section[] {Construction of Constant Crunch Surfaces}
\label{sec:1}

Over three decades ago Eardley and Smarr~\cite{Eardley1978}
carried out a generic classification of the
spacetimes that could be simulated numerically, and investigated
the limitations that the presence of singularities would impose. In
their paper they argued that CMC slicings
are particularly useful for numerical purposes. In particular, they
demonstrated this explicitly by constructing numerical
solutions to a wide array of dust collapse models. In a
similar vein, Brill et al.~\cite{Brill1978} explored the
nature of CMC slices of the Schwarzschild spacetime,
and they also presented some numerical examples.
In the present work, we 
explore the numerical utility of CMC slicings in the case of 
single black hole spacetimes.
For the sake of clarity, we will commence with a re-derivation
of the equations governing the CMC surfaces starting from 
Schwarzschild coordinates. From there, we will explore specific
properties of the surfaces, paying particular attention to
implications for numerical relativity.

We wish to find a spacelike hypersurface in the Schwarzschild
spacetime such that every point on the surface has the same constant
value of the trace of the extrinsic curvature tensor. 
We have at our
disposal the specification of the initial-value data, as
well as the freedom to choose the lapse and shift throughout
the evolution. To begin, let us take the standard
coordinate system of a single black hole spacetime of mass $M$
in Schwarzschild coordinates:
\begin{equation}
\label{Eq1}
ds^2 = -B(r) dt^2 + C(r) dr^2 + r^2 \left( d\theta^2 + 
\sin^2\theta\, d\varphi^2\right),
\end{equation}
with $B(r) = (1-2M/r)$ and $C(r)=1/B(r)$.  It will be convenient to
treat separately the regions inside and outside of the horizon. 
We will find that the two are related by an isometry.

\begin{figure}
\centerline{\epsfxsize=4.0truein\epsfbox{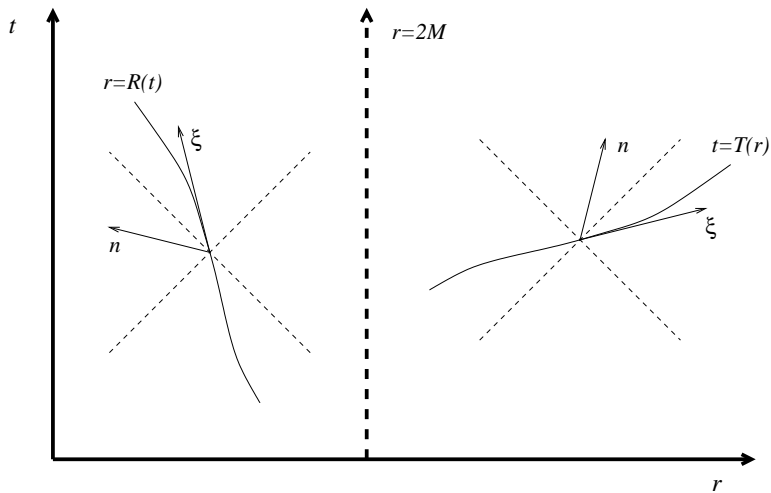}}
\caption{To construct CMC slices, at each
point we introduce  a unit normal vector ${\bf n}$ and the
unit tangent vector  \boldmath${\xi}$\unboldmath\ to the
surface. The local light cones are depicted by the 
light dashed lines.}
\label{Fig1}
\end{figure}

\subsection[] {Outside the Horizon ($r>2M$)}

Outside the horizon, CMC 
surfaces will be labeled by $t = T(r)$ 
(Fig.~\ref{Fig1}).  The requirement that the trace of the extrinsic
curvature be
constant throughout this surface yields a first order differential
equation for $T(r)$, determined by examining the behavior of the
normals to the surface.
The normal $\bf n$ to the spacelike hypersurface $T$ is given by,
\begin{equation}
{\bf n} = N_0 \nabla (t-T(r)) = n_t dt + n_r dr = N_0 (dt - T' dr),
\label{Eq2}
\end{equation}
where $N_0$ is a normalization constant and primes denote differentiation
with respect to $r$. The normalization is fixed by demanding that
\begin{eqnarray}
{\bf n} \cdot {\bf n} & = & -1 \\ 
& = & g^{tt} n_t n_t + g^{rr} n_r n_r \\
& = & N_0^2 \left(-\frac{1}{B} + \frac{1}{C} T'^2\right). 
\end{eqnarray}
Therefore,
\begin{equation}
N_0 = -\frac{1}{\sqrt{C-B T'^2}},
\end{equation}
and
\begin{eqnarray}
n_r & = & \frac{T'}{\sqrt{C-B T'^2}}, \\
n_t & = & -\frac{1}{\sqrt{C-B T'^2}}.
\end{eqnarray}
The contravariant components of the normal are given by
\begin{eqnarray}
n^r & = & g^{rr} n_r = \frac{T'}{C \sqrt{C-B T'^2}}, \\
n^t & = & g^{tt} n_t = \frac{1}{B\sqrt{C-B T'^2}}.
\end{eqnarray}
The trace of the extrinsic curvature is the fractional rate of contraction of 3-volume per
unit proper time along the normal, namely 
\begin{equation}
\label{Kequation}
K = -n^\alpha\,_{;\alpha} 
= -\frac{1}{r^2} \frac{d}{dr}\left(r^2 n^r\right)\,.
\label{eq:trk}
\end{equation}
Substitution of $n^r$ into Eq.~(\ref{eq:trk}) yields
a second-order ordinary differential equation for $T$.
Integrating this equation, we find
\begin{equation}
\label{TH}
H = \left( \frac{B r^2 T'}{\sqrt{C-BT'^2}} \right) + J,
\end{equation}
with $H$ an integration constant and $J$ an indefinite integral given
by
\begin{equation}
\label{J}
J = \int^r K \sqrt{BC} r^2 dr = \frac{1}{3} K r^3.
\end{equation}
Along the surface the rate of change of proper time, $d\tau$, with
proper distance, $ds$, is related to the slope of the surface $T$,
\begin{equation}
\frac{d\tau}{ds} = \sqrt{\frac{B}{C}}\ T'.
\end{equation} 
From Eq.~(\ref{TH}) we find
\begin{equation}
\label{Brill8}
\left( \frac{d\tau}{ds} \right)^2 = \frac{(H-J)^2}{(H-J)^2+Br^4}.
\end{equation}

\subsection[] {Inside the Horizon ($r<2M$)}

Finding the $K$-constant slices of Eq.~(\ref{Eq1}) within the horizon
is similar to the calculation done in the previous section; however,
as the roles of time and space coordinates reverse within the horizon,
we will find it useful to parameterize our spacelike surface as a
function of coordinate $t$, and look for $K$-constant surfaces of the
form (Fig.~\ref{Fig1})
\begin{equation}
r = R(t).
\end{equation}

The normal $\bf n$ to the spacelike hypersurface $R$ is given by
\begin{equation}
{\bf n} = N_0 \nabla (R(t)-r) = n_t dt + n_r dr = N_0 (\dot{R} dt - dr),
\end{equation}
with differentiation with respect to $t$ denoted by dots
and the $N_0$ a normalization constant fixed by:
\begin{eqnarray}
{\bf n} \cdot {\bf n} & = & -1 \\ 
& = & g^{tt} n_t n_t + g^{rr} n_r n_r \\
& = & N_0^2 \left(\frac{1}{C} - \frac{1}{B} \dot{R}^2\right). 
\end{eqnarray}
We have, therefore,
\begin{equation}
N_0 = \frac{-1}{\sqrt{C\dot{R}^2-B}},
\end{equation}
and
\begin{eqnarray}
n_r & = & \frac{1}{\sqrt{C\dot{R}^2-B}}, \\
n_t & = & \frac{-\dot{R}}{\sqrt{C\dot{R}^2-B}}.
\end{eqnarray}
The contravariant components of the normal are given by
\begin{eqnarray}
n^r & = & g^{rr} n_r = \frac{1}{C \sqrt{C\dot{R}^2-B}}, \\
n^t & = & g^{tt} n_t = \frac{\dot{R}}{B\sqrt{C\dot{R}^2-B}}.
\end{eqnarray}

From Eq.~(\ref{Kequation}), we once again find that fixing the trace
of the extrinsic curvature gives us a second-order differential
equation for $R(t)$, namely,
\begin{equation}
\label{Kinequation}
K = \frac{-2}{CR \sqrt{C\dot{R}^2-B}} -
\frac{2CB'\dot{R}^2-B\left(B'+C'\dot{R}^2
+2C\ddot{R}\right)}{2\left(C\dot{R}^2-B\right)^{3/2}},
\end{equation}
which can be simplified to
\begin{equation}
\label{R}
K\ R^2\dot{R} = - 
\frac{d}{dt}\left(\frac{BR^2}{\sqrt{C\dot{R}^2-B}}\right).
\end{equation}

Paralleling the approach from the last section, we introduce an
integration constant, $H$, and an indefinite integral $J$ (given by
Eq.~(\ref{J})), to obtain the first integral:
\begin{equation}
\label{RH}
H = \left(\frac{BR^2}{\sqrt{C\dot{R}^2-B}}\right)  + J.
\end{equation}
From Eq.~(\ref{R}) we find that the ``proper velocity'' along the
surface, $ds/d\tau = \sqrt{C/B} \dot{R}$, results in the same equation
both inside and outside of the horizon (Eq.~(\ref{Brill8})). This can
be rewritten as
\begin{equation}
\label{Brill11}
\left( \frac{ds}{d\tau} \right)^2 = 1\ +\ \frac{B\ R^4}{(H-J)^2}.
\end{equation}

The spacelike $K$-surfaces obtained from the first integrals,
Eqs.~(\ref{Brill8}) and (\ref{Brill11}), differ only by an isometry,
\begin{equation}
\label{isometry}
T' \Longleftrightarrow \frac{1}{\dot{R}}.
\end{equation}

\section[] {Properties of the $K$-surfaces}
\label{sec:3}

The spatial metric of a $K$-surface outside of the horizon is given by
\begin{eqnarray}
ds^2 & = & d{\ell}^2 + r^2 d\Omega^2 \\
     & = & \left(C-BT'^2\right) dr^2 + r^2 d\Omega^2.
\end{eqnarray}
Within the horizon it becomes
\begin{eqnarray}
ds^2 & = & d{\ell}^2 + r^2 d\Omega^2 \\
     & = & \left(C\dot R^2 -B\right) dt^2 + r^2 d\Omega^2.
\end{eqnarray}
These two expressions differ by the isometry of
Eq.~(\ref{isometry}). Using Eq.~(\ref{Brill8}) we can rewrite them in
terms of $H$ and $K$:
\begin{equation}
ds^2 = \frac{r^4}{\left(H-J\right)^2 + Br^4} dr^2 + r^2 d\Omega^2.
\label{K3metric}
\end{equation}
From this we arrive at the scalar curvature of the $K$-surface:
\begin{equation}
{}^{(3)}R = -\frac{2}{3} K^2\ +\ \frac{6H^2}{r^6}.   
\label{scalarR}
\end{equation}
Similarly, by using Eqs.~(\ref{Brill8}) and (\ref{Brill11}), the
extrinsic curvature associated with observers moving on world lines
orthogonal to the $K$-slices are also expressible in terms of $K$ and
$H$:
\begin{eqnarray}
\label{Krr}
K^{\hat{r}}_{\hat{r}} & = & \frac{1}{3} K\ +\ \frac{2H}{r^3}, \\
\label{Kpp}
K^{\hat{\theta}}_{\hat{\theta}} & = & K^{\hat{\phi}}_{\hat{\phi}} = 
\frac{1}{3} K\ -\  \frac{H}{r^3}.
\end{eqnarray}
The $K$-surfaces are therefore parametrized by two constants: the
trace of the extrinsic curvature tensor, $K$, and the constant of
integration, $H$. In addition one must fix a single point on the
surface, $t_o = T(r_o)$, which amounts to setting a time translation
parameter.  As can be seen in Eqs.~(\ref{scalarR})--(\ref{Kpp}), the
constant $H$ controls the variation of the intrinsic and extrinsic
curvatures over the $K$-surface.

\begin{figure}
\centerline{\epsfxsize=6.0truein\epsfbox{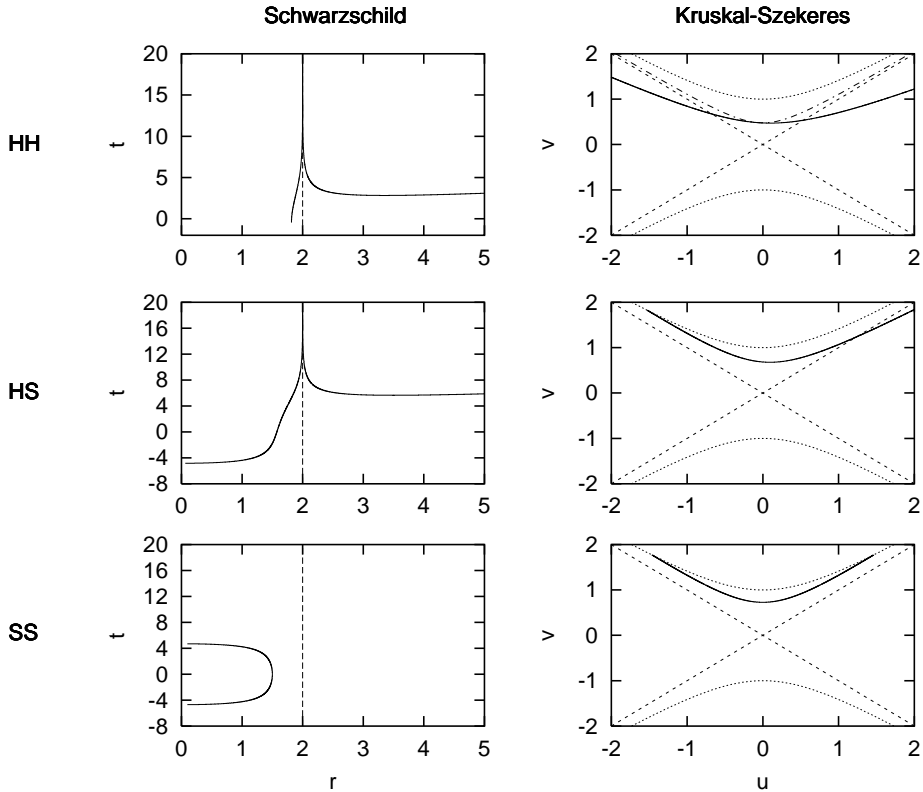}}
\caption{An example of the three families of spacelike $K$-surfaces
for $K=-0.1$.  The first row ({\em HH}\,) depicts a representative
horizon-to-horizon surface using $H=-1.25$, which corresponds to
$R_{min} \approx 1.816$ . This spacelike hypersurface is represented
both in Schwarzschild coordinates (left column) and in
Kruskal-Szekeres coordinates (right column). The middle
two graphs are the horizon-to-singularity ({\em HS}\,) surfaces using
$H = -1.43$. The bottom two graphs represents a typical
singularity-to-singularity ({\em SS}\,) surface. We have used $H=-1.25$ to
generate this {\em SS} $K$-surface.  The {\em HH} and {\em SS}
surfaces are close to their critical radii ($R_c \approx 1.5646$) and
therefore appear flattened, as described in the text.}
\label{SSHHHS}
\end{figure}

To elucidate the nature of the $K$-surfaces, we numerically integrate
Eqs.~(\ref{Brill8}) and (\ref{Brill11}). We find that within the
horizon there are 3 classes of $K$-constant surfaces, differentiated
by their asymptotic behavior. The singularity-singularity surfaces
({\em SS}\,) begin at the singularity aligned with the null surface, reach
up towards the horizon, and then fall back, reaching the singularity
along the null cone. The horizon-horizon surfaces ({\em HH}\,), which
we have also dubbed ``horizon-hugging'' surfaces, asymptote to the
horizon ($r\rightarrow2M$ for $|t|\rightarrow \infty$ in
Schwarzschild), dipping down towards the singularity in between.  This
feature was previously remarked upon by Brill {\em et al.}
\cite{Brill1978}. The asymptotes converge toward a null surface at the
horizon.  Finally, the horizon-singularity ({\em HS}\,) surfaces begin
at the horizon, and asymptote in to the singularity. Representative
surfaces for the value $K=-0.1$ are shown in Fig.~\ref{SSHHHS}. We
integrate the {\em HH} and {\em HS} surfaces across the
horizon into the region $r>2M$ by imposing continuity of the surface
and its first derivative at the horizon.  Because of the isometry,
Eq.~(\ref{isometry}), the surfaces outside of the horizon are
characteristically similar to those on the inside; in particular, both
sets are null at their asymptotes.

We have chosen to use the acronym {\em HH} for the horizon-to-horizon
hypersurfaces, in lieu of referring to them as ``regular''
hypersurfaces~\cite{Brill1978}, as each of the three types of
$K$-surfaces are, in a strict sense, regular.  In particular, each
surface asymptotes to a null surface, be it at the singularity or the
horizon. Observers on such a surface, or more precisely, observers
that are time synchronized throughout the surface, are never seen
crossing the horizon, nor do they ever reach the singularity! Outside
the horizon, every {\em HH} and {\em HS}
K-surface ($K\ne 0$) asymptotes for large
$r$ to null infinity. $K$-surfaces corresponding to positive values of
$K$ asymptote to past null infinity, and asymptote to future null infinity
for $K<0$.

\begin{figure}
\centerline{\epsfxsize=5.0truein\epsfbox{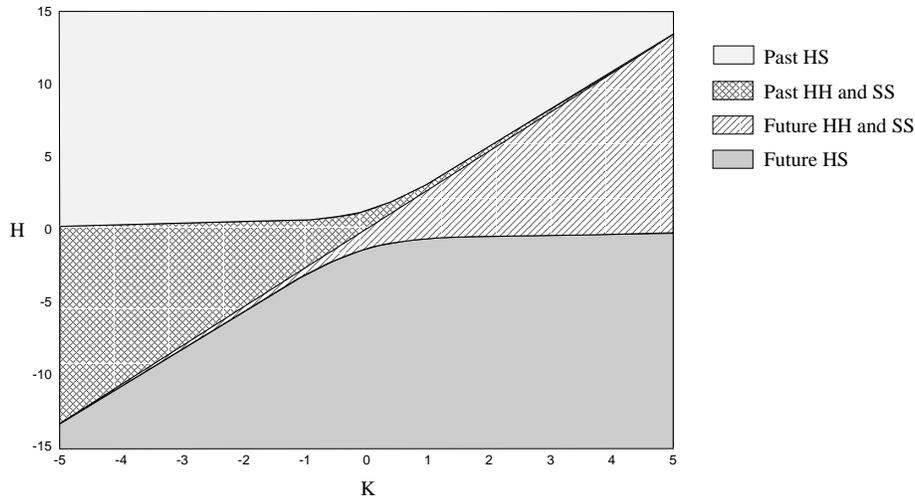}}
\caption{\protect\small {\em Phase space diagram of the classification of
$K$--surfaces.} The $K$--surfaces have been classified into
3 groups according to their asymptotic behavior (see
text). Which class a particular surface falls into depends
upon the values of $K$ and $H$, as detailed in the figure.
For the purposes of numerics, the surfaces of interest are
to be found in the ``Future {\em HH} and {\em SS}\/'' wedge, with $K<0$ (see
section~\ref{sec:3}).}
\label{contour}
\end{figure}

To gain a qualitative understanding of the $K$-constant foliation, it
is useful to analyze Eq.~(\ref{Brill11}) as an energy conservation
equation for a particle of unit total energy ($E=1$) moving in the
potential
\begin{equation}
\label{potential} 
V(r) = 
\frac{-\left(1-\frac{2M}{r}\right)r^4}{\left(H-\frac{1}{3}Kr^3\right)^2}.
\end{equation}
By using this energy equation we can determine the closest approach to
the singularity, $R_{min}$, of a given {\em HH} $K$-constant
surface. Two conditions must be satisfied to determine
$R_{min}$. First, the closest approach occurs when $\dot R =
(ds/d\tau) = 0$, which is equivalent to demanding
\begin{equation}
V(R_{min}) = 1.
\end{equation} 
This condition leads to a sixth-order polynomial in $R_{min}$:
\begin{equation}
\label{polynomial}
\frac{K^2}{9} R_{min}^6 + R_{min}^4 -
2\left(M+\frac{HK}{3}\right) R_{min}^3 + H^2 =0.
\end{equation}
Second, the solution for this surface at $R(t_{min})=R_{min}$ must be
concave, so as to rule out the {\em SS} surfaces (which bend towards
the singularity rather than the horizon). This is enforced by
demanding:
\begin{equation} 
\label{concave}
{\ddot R}(t_{min})|_{\dot R(t_{min})=0} = 
\frac{(R_{min}-2) \left[{-3+2R_{min}+KR_{min}^2\sqrt{\frac{2}{R_{min}}-1}}
\right]}{R_{min}^3}\geq 0.
\end{equation}
The solution of these two conditions, as shown in
Fig.~\ref{contour}, gives rise to the emergence of two
critical values for $H$, namely $H_\pm$, for a given value
of $K$. In addition, an {\em HH} surface can be made to approach
arbitrarily closely to the singularity at $r=0$ by choosing
an appropriately large positive value of $K$.  Negative
values of $K$ tend to ``hug the horizon.''

For fixed $K$ and $H$ we know how to compute how closely a
CMC surface comes to the singularity. But, for a given value
of $K$, what value of $H$ gives the closest overall approach
to the singularity?  We can determine the critical values
for $R$ and $H$, given by $R_c$ and $H_{\pm}$ respectively,
by looking at the point where the first and second time
derivatives of $R(t)$ vanish. $H_-$ occurs along the lower
boundary of the contour plot in Fig.~\ref{contour}, and
corresponds to the $K$-surface that reaches down the furthest
towards the singularity for a given value of $K$. The
vanishing of $\ddot R|_{\dot R=0}$ leads to the following
equation for $R_c$:
\begin{equation}
\label{rcriteqn} 
\left(R_c-2\right) \left(-3+2R_c+KR_c^2\sqrt{\frac{2}{R_c}-1}\right) = 0,
\end{equation}
which can be rewritten as a 4th order polynomial in $R_c$
\begin{equation}
\label{4poly}
K^2 R_c^4 - 2K^2R_c^3+4R_c^2-12R_c+9=0.
\end{equation}
One must take care in examining the roots of this equation, as there
are more solutions to Eq.~(\ref{4poly}) than there are for
Eq.~(\ref{rcriteqn}).  Nevertheless this equation gives two distinct
real roots, depending on the sign of $K$:
\begin{eqnarray}
\label{rcritp}
R_c|_{K>0} & = & \frac{1}{2} - \frac{1}{2} 
   \sqrt{3-\frac{8}{K^2}-\chi+\frac{2}{\sqrt{\chi}}+\frac{16}{K^2 \sqrt{\chi}}}
   + \frac{\sqrt{\chi}}{2}, \\
\label{rcritn}
R_c|_{K<0} & = & \frac{1}{2} + \frac{1}{2} 
   \sqrt{3-\frac{8}{K^2}-\chi+\frac{2}{\sqrt{\chi}}+\frac{16}{K^2 \sqrt{\chi}}}
   + \frac{\sqrt{\chi}}{2},
\end{eqnarray}
where
\begin{eqnarray}
\label{xi} 
\xi & \equiv & 32+108K^2+243K^4+27K^2\sqrt{16+56K^2+81K^4}, \\
\label{chi}
\chi & \equiv & 1 - \frac{8}{3K^2} + \frac{16+36K^2}{3\ 2^{1/3} K^2 {\xi}^{1/3}}
     + \frac{2^{1/3} {\xi}^{1/3}}{3 K^2}.
\end{eqnarray}
When $K=0$ we see from Eq.~(\ref{rcriteqn}) that $R_c=3/2$. The
regular $K$-surfaces are thus bounded between the horizon at 
$R_+=2 M$
and $R_-=R_c M$.  Using Eq.~(\ref{RH}), and setting $\dot R=0$, we
obtain, for the case of a black hole in Schwarzschild coordinates
\begin{equation} 
\label{Hpm}
H_{\pm} = \frac{B_{\pm} R_{\pm}^2}{\sqrt{-B_{\pm}}} +
\frac{K R_{\pm}^3}{3},
\end{equation}
with
\begin{equation}
B_{\pm} = \left(1-\frac{2M}{R_{\pm}}\right).
\end{equation}
We therefore have
\begin{equation}
\label{Hp}
H_+ = \frac{8}{3} M^3 K.
\end{equation}

\begin{figure}[t]
\centerline{\epsfxsize=3.0truein\epsfbox{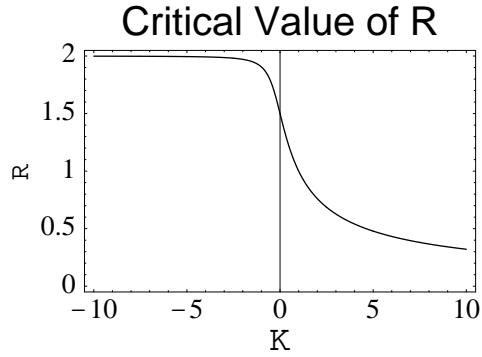}}
\caption{ {\em Critical value of $R_c=R_-$ as a function of $K$ for
Schwarzschild coordinates and $M=1$.} One can readily see 
that the $K$-surface ``hugs'' the horizon for large negative
values of $K$.}
\label{rhcrit}
\end{figure}

For large values of $|K|$ one can show that the critical value of $R$,
$R_-$, depends upon the sign of $K$. In particular,
\begin{equation} 
R_c \longrightarrow  \left\{    \begin{array}{ll}
        2-\frac{1}{8} K^{-2} & \mbox{for $K \ll -1$} \\
        \left(\frac{9}{4}\right)^{1/3}\ \ K^{-2/3} & \mbox{for $K \gg 1$}
                \end{array}
                \right..
\end{equation}
This in turn gives the following asymptotic values for $H_{m}$:
\begin{equation} 
H_- \longrightarrow  \left\{    \begin{array}{ll}
        -\frac{8}{3}K\ -\ \frac{1}{2K} & \mbox{for $K \ll -1$} \\
        \frac{-9}{2K} & \mbox{for $K \gg 1$}
                \end{array}
                \right..
\end{equation}

\section[] {Constant Crunch Coordinates:  A Spacetime Metric
for a $K$-surface Foliation of the Schwarzschild Black
Hole.}
\label{sec:4}

A number of features of the $K$-constant surfaces presented in the
previous sections seem particularly well suited to the numerical
analysis of generic black hole spacetimes.  First, the surfaces
asymptote to a null surface, making them effective
for gravity wave extraction.  Second, the $K$-surfaces naturally avoid
the crushing singularity.  Finally, for large negative values of $K$
the surfaces ``hug the horizon.'' This last feature, illustrated in
Fig.~\ref{rhcrit}, allows one to focus attention on the region
relevant for gravity wave generation---the region outside the horizon.

\begin{figure}
\centerline{\epsfxsize=5.0truein\epsfbox{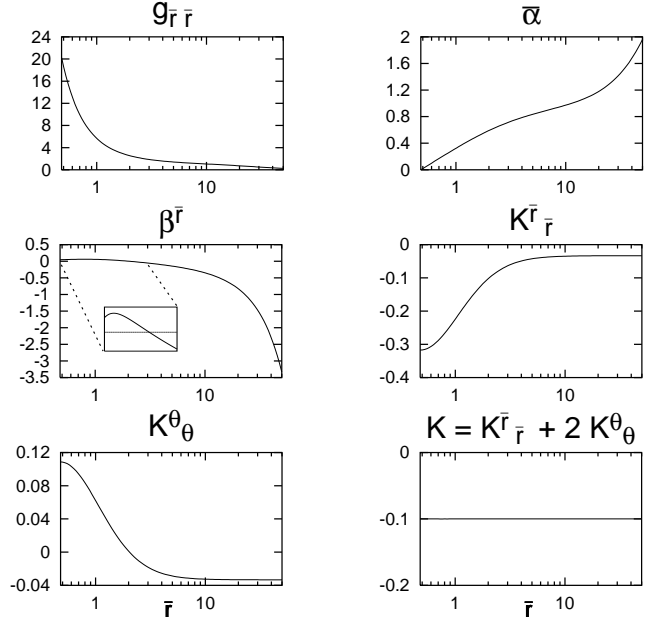}}
\caption{{\em The radial behavior of the various components of the
spacetime metric (Eq.~(\ref{Kmetric})), under the isotropic-like
transformation given in Eq.~(\ref{iso}), for $M=1$, $K=-0.1$,
$H=-1.0$ and $\bar{R}_{min}\approx 0.47922$.}  From upper left to lower
right we plot the radial metric component ($g_{\bar{r}\bar{r}}$),
lapse ($\bar{\alpha}$), radial shift ($\beta^{\bar{r}}$), diagonal
components of the extrinsic curvature tensor ($K^{\bar{r}}_{\bar{r}}$
and $K^{\bar{\theta}}_{\bar{\theta}}= K^{\bar{\phi}}_{\bar{\phi}}$),
and in the lower right frame the trace of the extrinsic curvature
tensor ($K = K^{\bar{r}}_{\bar{r}} + K^{\bar{\theta}}_{\bar{\theta}} +
K^{\bar{\phi}}_{\bar{\phi}}$) as a consistency check. All of the
functions are regular and well behaved. The growth of the lapse and
shift for large $\bar{r}$ is expected, as the surfaces become
asymptotically null.  We show an exploded view of the behavior of the 
radial shift near the throat to emphasize that the shift changes 
sign and becomes positive before reaching $\bar{r}=\bar{R}_{min}$.}
\label{components}
\end{figure}

In this section we generate a $K$-constant foliation for the
Schwarzschild black hole that, in addition to the properties just
mentioned, also has regular and static metric and extrinsic curvature
components.  To generate this $K$-constant slicing we use the
coordinate transformation
\begin{eqnarray}
\bar t & = & t - T(r),\\
\rho   & = & r.
\end{eqnarray}
Under this transformation, the metric from Eq.~(\ref{Eq1}) becomes
\begin{equation}
\label{Kmetric}
ds^2 = -(1-\frac{2M}{\rho}) d{\bar t}^2 
       + 2 \frac{(J-H)}{\sqrt{(H-J)^2 
       + (1-\frac{2M}{\rho}) \rho^4}} d{\bar t} d\rho
       + \frac{\rho^4}{(H-J)^2 + (1-\frac{2M}{\rho}) \rho^4} d\rho^2
       + \rho^2 d\Omega^2.
\end{equation}
The constant $\bar t$ slices of this metric are $K$-constant
surfaces. It is to be noted that Eq.~(\ref{Kmetric}) agrees with
Eq.~(53) of Iriondo et al. (for the case of constant $K$ and
vacuum).~\cite{Iriondo1996} However, in order to regularize the
$g_{\rho\rho}$ metric component at the throat, we add the
isotropic-like radial transformation~\cite{Matzner2000}:
\begin{equation}
\bar r=
{1\over2}\left({-{1\over2}R_{min}+\rho+\sqrt{-R_{min}\rho+\rho^2}}\right),
\label{iso}
\end{equation} 
with $R_{min}$ the minimum coordinate location of the throat, given by
Eq.~(\ref{polynomial}).  This coordinate representation of a black
hole spacetime provides a foliation with $K$-constant, $H$-constant
spacelike hypersurfaces. Each hypersurface is metrically equivalent to
all others---the surfaces are independent of $\bar t$, and hence
static. In addition, the hypersurfaces are asymptotically null
($T'(r)\rightarrow 1$ as $r\rightarrow\infty$).  Furthermore, the
lapse, the shift, and all of the 3-metric and extrinsic curvature
components are regular and well behaved, as illustrated in
Fig.~\ref{components}.

In addition to restricting ourselves to the
singularity-avoiding family ({\em HH}) of $K$-surfaces, two
additional conditions on the $K$-surfaces are demanded by
the nature of our problem -- the eventual simulation of the
gravity-wave emission from two interacting black
holes. First, the foliations must asymptote at large $\bar r$
to future null infinity. Therefore, we must restrict our
attention to negative values for the trace of the extrinsic
curvature tensor, $K$. Second, we require that such negative
$K$ hypersurfaces enter the future singularity region. This
will ensure proper coverage of the relevant region just
above the future horizon, which is precisely where the
gravity waves are produced. However, we expect the
initial-data formulation for such surfaces to be involved,
and this may guide our choices even more systemically.  
The two additional requirements  limit us to
the relatively narrow wedge of Fig.~\ref{contour}, formed
by restricting to the ``Future {\em HH} and {\em SS}\,'' shaded region
with $K<0$.

\begin{figure}
\centerline{\epsfxsize=5.0truein\epsfbox{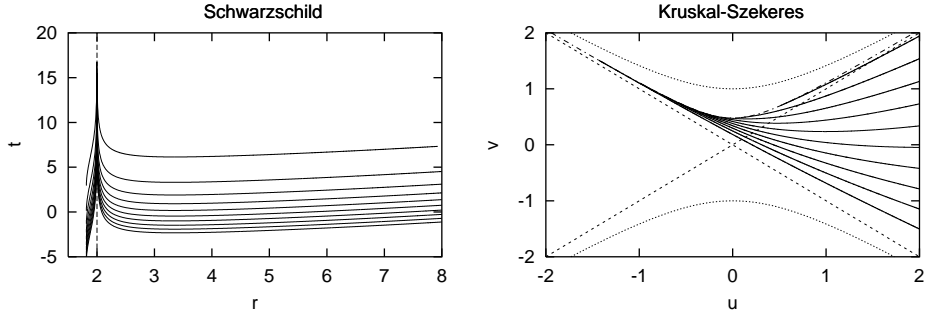}}
\caption{{\em The constant crunch coordinate foliation of the
Schwarzschild spacetime.} We show the foliation generated by
Eq.~(\ref{Kmetric}) for $M=1$, $K=-0.1$, $H=-1.25$ and
$R_{min}\approx 1.816$, in Schwarzschild and Kruskal-Szekeres
coordinates.}
\label{foliate}
\end{figure}

\begin{figure}
\centerline{\epsfxsize=5.0truein\epsfbox{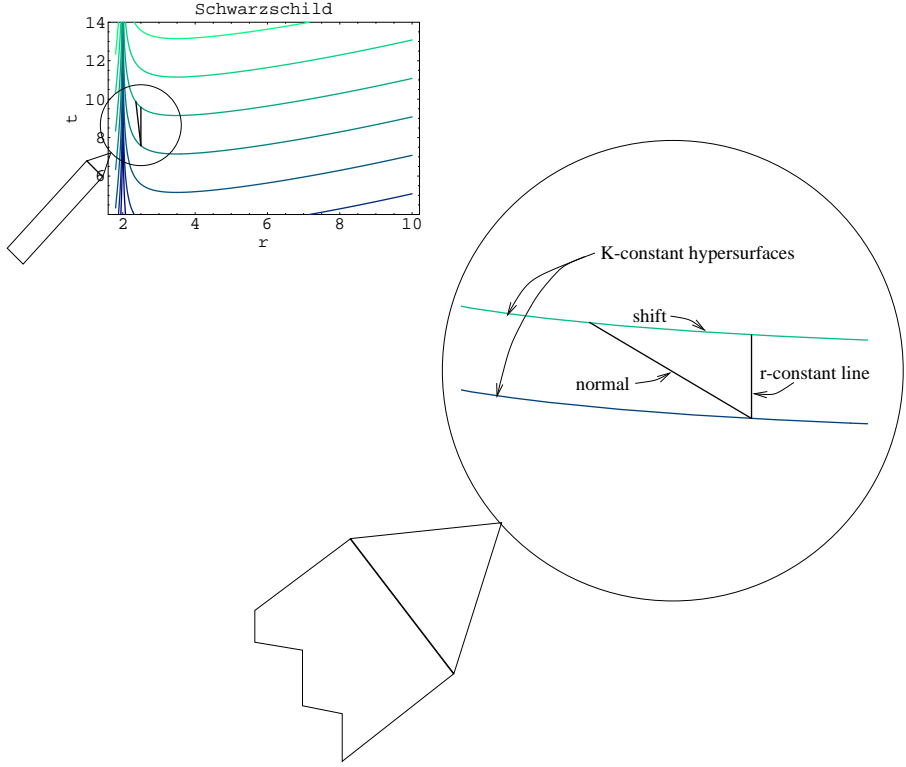}}
\caption{{\em The non-zero shift in the constant-crunch
coordinatization of the Schwarzschild spacetime.} We show a foliation
generated by Eq.~(\ref{Kmetric}) for $M=1$, $K=-1$, $H\approx-3.11$,
and $R_{min}\approx 1.9$. We show the $r=\mbox{\em constant}$ line
segment at a point on one of the surfaces, together with the
inward-pointing normal vector.  The misalignment of these two line
segments indicates the non-zero shift in this static coordinatization.
A magnified view of the two line segments is shown in the bottom
panel.}
\label{shift}
\end{figure}

A representative constant-crunch foliation generated by Eq.~(\ref{Kmetric}) is
shown in Fig.~\ref{foliate}.  The avoidance of ``grid stretching'' is
accomplished by a suitable choice of shift vector.  To illustrate the
non-zero shift we show the $K=-1$, $H\approx-3.11$ foliation of
Schwarzschild in Fig.~\ref{shift}, with the explicit misalignment of
the $r=\mbox{\em constant}$ line segment and the normal vector.

\begin{figure}
  \begin{center} \epsfig{file=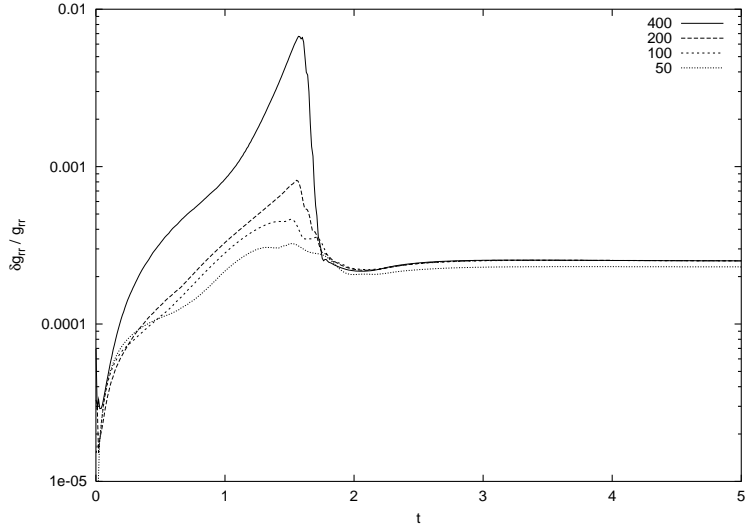, width=4in}
  \end{center} \caption{{\em A numerical example:
  convergence of the radial metric component for the
  evolution of an HH surface spanning the past singularity
  region and reaching positive null infinity.} We show the
  time evolution of a single black hole with $M=1, K=-1,
  H=-3$, and plot the mean fractional error in the radial
  metric component $\bar a=\sqrt{g_{\bar r \bar
  r}}$. Results are shown for four different resolutions,
  $N=400/2^p$ ($p=0,1,2,3$), where each curve has been
  rescaled by a factor of $4^p$.  For $t>2$ the solution
  displays approximately second order convergence, with a
  fractional error of around $0.1\%$ for $N=200$.  Long term
  evolution is stable, and the system has been succesfully evolved
  beyond $t=50,000M$.}
\label{convergence}
\end{figure}

\begin{figure}
  \begin{center} \epsfig{file=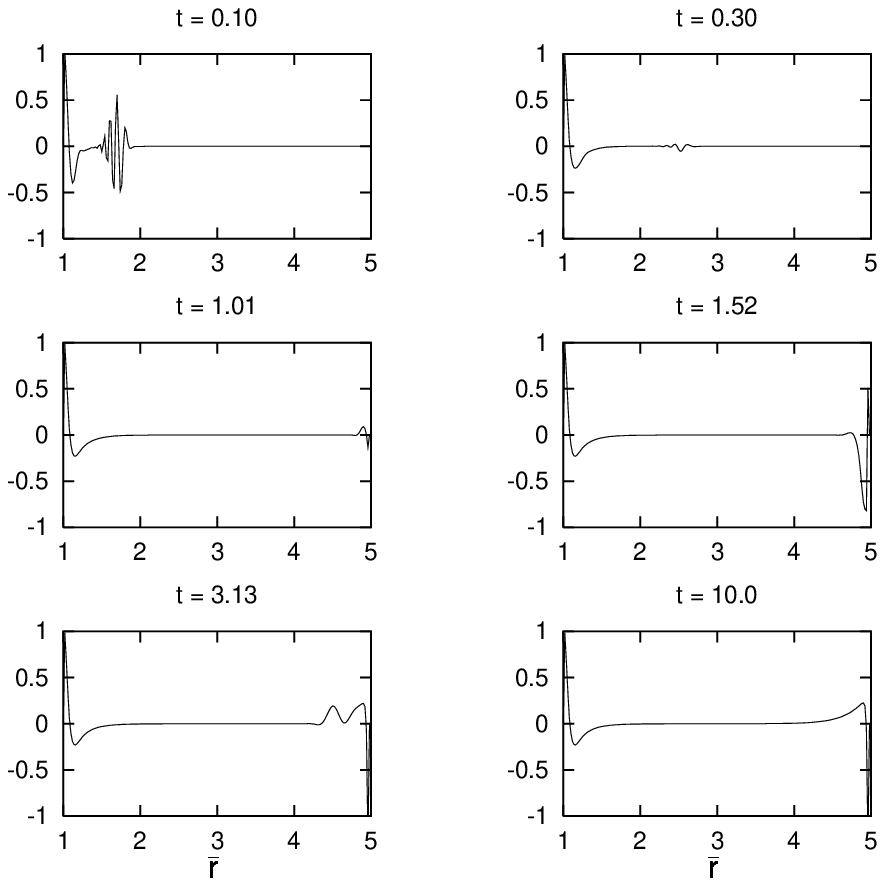, width=6in} \end{center}
    \caption{{\em Snapshots of the Hamiltonian constraint.}  For the
    same parameter choices as in Fig.~\ref{convergence} we display the
    Hamiltonian constraint during the early stages of evolution.  A
    wave generated by truncation error at the inner boundary
    propogates outwards, is reflected by the outer boundary back into
    the domain, and then rapidly decays. No further evolution is seen
    in the Hamiltonian constraint after $t\approx 5$.}
\label{hamiltonian}
\end{figure}

Finally, we describe several preliminary numerical experiments using
$K$-surfaces.  A full treatment of a single black hole using
$K$-constant foliations will be presented elsewhere
\cite{Shoemaker2000}. Here we present several
sample evolutions demonstrating the utility of constant mean curvature
slicings.  Figures \ref{convergence} and \ref{hamiltonian} display
results from the simplest possible test of $K$-constant foliation of
the Schwarzschild geometry; the domain is taken to be a thin shell
close to the horizon (in this case $r \in [1,5]$), analytic Dirichlet
conditions are applied at both boundaries of the computational domain,
and analytic lapse and shift conditions obtained from
Eqs. (\ref{Kmetric}) and (\ref{iso}) are used. The figures represent
the singularity-avoiding foliation $K=-1$ and $H=-3$, for which the
evolution was found to be stable and accurate over very long
timescales.  Using $50$ grid points, the code succesfully ran beyond
$t=50,000M$ while maintaining high accuracy. The fractional error in
the metric components was typically $1-2\%$.

Fig.~\ref{convergence} shows the convergence of the mean fractional
error in the metric component $\bar a=\sqrt{g_{\bar r\bar r}}$ as a
function of time.  Each curve has been rescaled by a factor of $4^p$,
where the number of grid points is given by $400/2^p$, with
$p=0,1,2,3$.  For $t>2$ the solution is approximately second
order accurate.  Initially, noise generated on the inner boundary
causes fluctuations whose magnitude is largely independent of the
number of grid points.  Fig.~\ref{hamiltonian} shows snapshots of
the Hamiltonian constraint at various times in the evolution.
Qualitatively, a wave which is triggered by truncation error is seen
to propagate outwards from the inner boundary.  The amplitude reduces
rapidly, before growing once more as it is reflected off the outer
boundary.  The magnitude and speed of propagation of the wave quickly
decay as the wave moves back into the domain, leaving a static
solution which is stable beyond $t=50,000M$.  The numerical error
which initially propagates through the domain is caused entirely by
the analytic Dirichlet boundary conditions, and can be largely
eliminated by the use of more realistic conditions.
\cite{Shoemaker2000}

The numerical runs presented here evolve a $K=-1$
singularity-avoiding hypersurface that asymptotes to null infinity,
entering the past singularity region. This is not of the
class of $K$-surfaces emphasized in this paper for use in
numerical relativity.  Ideally, we would have
preferred presenting the evolution of a $K=-1$ {\em HH} surface that
enters the future singularity region. However, we were unable to
find an {\em HH} surface in the future singularity region that evolved
stably using the naive Dirichlet boundary conditions and analytic
lapse and shift conditions. Nevertheless, we have evolved such
surfaces stably by incorporating (1) area locking shift conditions,
(2) isometery conditions at the throat, and (3) an outgoing boundary
condition based on the difference between the numeric and analytic
solutions. These numerical results, and the corresponding stability
analysis, are not within the scope of this paper and will be presented
elsewhere.\cite{Shoemaker2000}

\section[] {From One Black Hole to Two Spinning Black Holes}
\label{sec:5}

In this paper we considered a static spacetime metric for the
Schwarzschild black hole, with spacelike hypersurfaces of constant
(not necessarily zero) value of the trace of the extrinsic curvature
tensor. This slicing provides a natural generalization of the maximal
slicing scheme currently in use in some numerical approaches to the
binary black hole problem.

An essential feature of our $K$-constant metric is a spatially-varying
radial shift vector, which allows the surfaces to avoid the
singularity while evading the grid stretching problems often
encountered with other metrics. Work is in progress to develop a
geometric handle on our shift condition, akin to that of the ``minimal
distortion'' shift often discussed.  The inner and outer boundary
conditions are also particularly convenient.  In the inner regions,
the ``horizon hugging'' feature of the $K$-slices, together with their
regularity, may remove the need to excise the grid within the apparent
horizon, thus providing a natural ``boundary without a boundary''
avoidance of the singularity. In addition, at the outer boundary the
surfaces are asymptotically null, which may aid in gravity wave
extraction.

In examining the characteristics of the $K$-slices we reviewed the
three families of $K$-surfaces, including the horizon-to-singularity
({\em HS}\,) surfaces, as well as the more familiar horizon-to-horizon
({\em HH}\,) and singularity-to-singularity ({\em SS}\,) surfaces. All
three families of surfaces either asymptote to the singularity, or to
spatial infinity along a null surface.  Thus the family of observers
that make up such a foliation never ``observe'' the surfaces reach the
singularity. This suggests a possible payoff in using the {\em HS}
surfaces in the evolution of black hole spacetimes, as there is no
danger of the surface reaching the singularity. Just as there are
natural boundary conditions which ``freeze out'' gravitational
radiation as it progresses to null infinity, so too can we expect
boundary conditions where the dynamic space freezes out as it
approaches the singularity. Are there such natural boundary conditions
at the future singularity? It would be worthwhile to investigate such
asymptotically-null boundary conditions at the singularity of an $HS$
surface.

The future utility of numerical simulations of black hole spacetimes
hinges in large part on a suitable choice of coordinates for the
initial data, and on the particular evolution of these coordinates
through the four lapse and shift conditions.  These conditions are the
only handles by which to manage the growth of the metric and curvature
components during evolution.  It is through judicious choices of
lapse and shift that one is able to effectively enable singularity
avoidance, and allow for efficient extraction of gravitational
radiation. The simplistic example presented here may provide some
guidance as to how to proceed in the more general case. $K$-surfaces
in the generic two black hole problem can be expected to preserve the
singularity-avoidance ``horizon hugging'' behavior, as well as remain
asymptotically null at the boundaries. It remains to be seen what
shift vectors are required to manage the growth of the intrinsic and
extrinsic properties of the metric in the more general case, though
the constant-crunch shift presented here is a natural starting point.

We are currently pursuing three avenues towards furthering and
generalizing the work presented here. First, we are analyzing the
stability of this coordinatization to small perturbations. Second, we
are using the (1+1)-dimensional code discussed in the last section to
examine a greater portion of the parameter space of initial data, so
as to fully explore the numerical stability of the slicing.  From our
preliminary numerical investigations, we expect to be able to do a
full spacetime ``evolution'' of Schwarzschild, and have it run stably
and accurately for extended periods of time over a wide range of
$H$-$K$ parameter space.  Finally, we are analyzing the
Oppenheimer-Snyder collapse in such $K$-slicings, following the lead
of Eardley and Smarr,~\cite{Eardley1978} as this will further test our
slicing in a non static setting.~\cite{Thorne2000}

\begin{acknowledgements}

We wish to acknowledge the Los Alamos National Laboratory
LDRD/ER program for financial support. WAM wishes to thank
the Institute for Theoretical Physics at UCSB for providing
a stimulating working environment in which to complete part
of this research.  DH was supported in part by the National
Science Foundation under Grant No. PHY9907949 to the ITP.
PL was supported in part by NSF grants PHY9800973 and
PHY9800970.  We wish to thank Richard Matzner for advice on
handling the coordinate ambiguity at the throat.  We are
especially grateful to John A. Wheeler for encouraging us to
examine these slices in the context of the numerical
treatment of black hole spacetimes.
\end{acknowledgements}

\end{document}